\PassOptionsToPackage{unicode}{hyperref}
\PassOptionsToPackage{hyphens}{url}
\PassOptionsToPackage{dvipsnames,svgnames,x11names}{xcolor}
\documentclass[
]{article}
\usepackage{amsmath,amssymb}
\usepackage{lmodern}
\usepackage{iftex}
\ifPDFTeX
  \usepackage[T1]{fontenc}
  \usepackage[utf8]{inputenc}
  \usepackage{textcomp} % provide euro and other symbols
\else % if luatex or xetex
  \usepackage{unicode-math}
  \defaultfontfeatures{Scale=MatchLowercase}
  \defaultfontfeatures[\rmfamily]{Ligatures=TeX,Scale=1}
\fi
\IfFileExists{upquote.sty}{\usepackage{upquote}}{}
\IfFileExists{microtype.sty}{% use microtype if available
  \usepackage[]{microtype}
  \UseMicrotypeSet[protrusion]{basicmath} % disable protrusion for tt fonts
}{}
\makeatletter
\@ifundefined{KOMAClassName}{% if non-KOMA class
  \IfFileExists{parskip.sty}{%
    \usepackage{parskip}
  }{% else
    \setlength{\parindent}{0pt}
    \setlength{\parskip}{6pt plus 2pt minus 1pt}}
}{% if KOMA class
  \KOMAoptions{parskip=half}}
\makeatother
\usepackage{xcolor}
\NewDocumentCommand\citeproctext{}{}
\NewDocumentCommand\citeproc{mm}{%
  \begingroup\def\citeproctext{#2}\cite{#1}\endgroup}
\makeatletter
 \let\@cite@ofmt\@firstofone
 \def\@biblabel#1{}
 \def\@cite#1#2{{#1\if@tempswa , #2\fi}}
\makeatother
\newlength{\cslhangindent}
\newlength{\csllabelwidth}
\newenvironment{CSLReferences}[2] % #1 hanging-indent, #2 entry-spacing
 {\begin{list}{}{%
  \setlength{\itemindent}{0pt}
  \setlength{\leftmargin}{0pt}
  \setlength{\parsep}{0pt}
  \ifodd #1
   \setlength{\leftmargin}{\cslhangindent}
   \setlength{\itemindent}{-1\cslhangindent}
  \fi
  \setlength{\itemsep}{#2\baselineskip}}}
 {\end{list}}
\usepackage{calc}

\ifLuaTeX
\usepackage[bidi=basic]{babel}
\else
\usepackage[bidi=default]{babel}
\fi
\babelprovide[main,import]{american}
\def\languageshorthands#1{}
\ifLuaTeX
  \usepackage{selnolig}  % disable illegal ligatures
\fi
\IfFileExists{bookmark.sty}{\usepackage{bookmark}}{\usepackage{hyperref}}
\IfFileExists{xurl.sty}{\usepackage{xurl}}{} % add URL line breaks if available
\hypersetup{
  pdftitle={AMReX-Astrophysics Microphysics: A set of microphysics
routines for astrophysical simulation codes based on the AMReX library},
  pdfauthor={AMReX-Astro Microphysics Team, Khanak Bhargava, Abigail
Bishop, Zhi Chen, Doreen Fan, Carl Fields, Adam M. Jacobs, Eric T.
Johnson, Max P. Katz, Mark Krumholz, Chris Malone, Andy Nonaka, Piyush
Sharda, Alexander Smith Clark, Frank Timmes, Ben Wibking, Don E.
Willcox, Michael Zingale},
  pdflang={en-US},
  colorlinks=true,
  linkcolor={Maroon},
  filecolor={Maroon},
  citecolor={Blue},
  urlcolor={Blue},
  pdfcreator={LaTeX via pandoc}}

\title{AMReX-Astrophysics Microphysics: A set of microphysics routines
for astrophysical simulation codes based on the AMReX library}

\definecolor{c53baa1}{RGB}{83,186,161}
\definecolor{c202826}{RGB}{32,40,38}

\usepackage[affil-it]{authblk}
\usepackage{orcidlink}
\author[1%
  ]{AMReX-Astro Microphysics Team%
    }
\author[2%
  ]{Khanak Bhargava%
    \,\orcidlink{0000-0003-0385-7918}\,%
    }
\author[3%
  ]{Abigail Bishop%
    \,\orcidlink{0000-0002-0375-644X}\,%
    }
\author[2%
  ]{Zhi Chen%
    \,\orcidlink{0000-0002-2839-107X}\,%
    }
\author[4%
  ]{Doreen Fan%
    \,\orcidlink{0000-0002-3246-4315}\,%
    }
\author[5%
  ]{Carl Fields%
    \,\orcidlink{0000-0002-8925-057X}\,%
    }
\author[4%
  ]{Adam M. Jacobs%
    \,\orcidlink{0000-0002-3580-2420}\,%
    }
\author[2%
  ]{Eric T. Johnson%
    \,\orcidlink{0000-0003-3603-6868}\,%
    }
\author[2%
  ]{Max P. Katz%
    \,\orcidlink{0000-0003-0439-4556}\,%
    }
\author[6%
  ]{Mark Krumholz%
    \,\orcidlink{0000-0003-3893-854X}\,%
    }
\author[7%
  ]{Chris Malone%
    \,\orcidlink{0000-0002-4045-7932}\,%
    }
\author[8%
  ]{Andy Nonaka%
    \,\orcidlink{0000-0003-1791-0265}\,%
    }
\author[9%
  ]{Piyush Sharda%
    \,\orcidlink{0000-0003-3347-7094}\,%
    }
\author[2%
  ]{Alexander Smith Clark%
    \,\orcidlink{0000-0001-5961-1680}\,%
    }
\author[10%
  ]{Frank Timmes%
    \,\orcidlink{0000-0002-0474-159X}\,%
    }
\author[11%
  ]{Ben Wibking%
    \,\orcidlink{0000-0003-3175-2291}\,%
    }
\author[12%
  ]{Don E. Willcox%
    \,\orcidlink{0000-0003-2300-5165}\,%
    }
\author[2%
  ]{Michael Zingale%
    \,\orcidlink{0000-0001-8401-030X}\,%
    }

\affil[1]{Collaboration%
  }
\affil[2]{Department of Physics and Astronomy, Stony Brook University,
Stony Brook, NY, USA%
  }
\affil[3]{Department of Physics, University of Wisconsin, Madison,
Madison, WI, USA%
  }
\affil[4]{Independent Researcher, USA%
  }
\affil[5]{Department of Astronomy, University of Arizona, Tucson, AZ,
USA%
  }
\affil[6]{Research School of Astronomy and Astrophysics, The Australian
National University, Australia%
  }
\affil[7]{Los Alamos National Laboratory, Los Alamos, NM, USA%
  }
\affil[8]{Lawrence Berkeley National Laboratory, Berkeley, CA, USA%
  }
\affil[9]{Leiden Observatory, Leiden, The Netherlands%
  }
\affil[10]{Arizona State University, Tempe, AZ, USA%
  }
\affil[11]{Department of Physics and Astronomy, Michigan State
University, E. Lansing, MI, USA%
  }
\affil[12]{Institute for Advanced Computational Science, Stony Brook
University, Stony Brook, NY, USA%
  }
\date{4 Aug 2026}

\begin{document}
\maketitle

\section{Summary}\label{summary}

The AMReX-Astrophysics Microphysics library provides a common set of
microphysics routines (reaction networks and associated physics,
equations of state, and various transport coefficients) as well as
solvers (stiff ODE integrators, nonlinear system solvers) for
astrophysical simulation codes built around the AMReX adaptive mesh
refinement library (\citeproc{ref-amrex}{W. Zhang et al., 2019}).
Several multi-dimensional simulation codes, including the compressible
hydrodynamics code Castro (\citeproc{ref-castro_I}{Almgren et al.,
2010}), the low-Mach number hydrodynamics code MAESTROeX
(\citeproc{ref-maestroex}{Fan et al., 2019}), and the
radiation-hydrodynamics code Quokka (\citeproc{ref-quokka}{Wibking \&
Krumholz, 2022}) use Microphysics to provide the physics and solvers
needed to close the hydrodynamics systems that they evolve. The library
is implemented in C++ with GPU-offloading a key design feature.

\section{Statement of need}\label{statement-of-need}

Astrophysical simulation codes need many different small-scale
(microphysics) physics inputs to close the system of equations. There
are many astrophysics simulation codes built around the AMReX library,
with each specializing in different astrophysics phenomena. Each of
these codes share some common needs. The Microphysics library was
created to minimize developer effort across these codes and coordinate
the approach to exascale compute architectures, in particular, GPU
support for astrophysical simulation codes.

\section{State of the field}\label{state-of-the-field}

Individual reaction networks and equations of state have been made
available by authors for decades, including a wide variety from
\href{https://cococubed.com/code_pages/burn.shtml}{Cococubed.com}. Flash
(\citeproc{ref-flash}{Fryxell et al., 2000}) comes with a set of
reaction networks and equations of state as well. The closest
compilation to ours is the recent singularity-EOS library
(\citeproc{ref-singularity}{Miller et al., 2024}), which provides
various equations of state. Microphysics predates this library, going
back to at least 2013. Further, our design targets codes that use the
AMReX library directly, and encompasses a wider set of physics.

\section{Software design}\label{software-design}

The Microphysics project started in 2013 as a way to centralize the
reaction networks and equations of state used by Castro and MAESTRO
(\citeproc{ref-maestro}{Nonaka et al., 2010}), the predecessor to
MAESTROeX. Originally, Microphysics used Fortran and for a brief period,
it was referred to as Starkiller Microphysics, which was an attempt to
co-develop microphysics routines for the Castro and the Flash
(\citeproc{ref-flash}{Fryxell et al., 2000}) simulation codes. As
interest in GPUs grew (with early support added to Microphysics in
2015), Castro moved from a mix of C++ and Fortran to pure C++ to take
advantage of GPU-offloading afforded by the AMReX library, and C++ ports
of all physics routines and solvers were added to Microphysics. At this
point, the project was formally named the AMReX-Astrophysics
Microphysics library. Today, the library is completely written in C++
and relies heavily on the AMReX data structures to take advantage of
GPUs. The GPU-enabled reaction network integrators led to the Quokka
code adopting Microphysics for their simulations.

Microphysics provides several different types of physics: equations of
state, reaction networks and screening methods, nuclear statistical
equilibrium solvers and tabulations, thermal conductivities, and
opacities, as well as the tools needed to work with them, most notably
the suite of stiff ODE integrators for the networks. Several classic
Fortran libraries have been converted to header-only C++
implementations, including the VODE integrator
(\citeproc{ref-vode}{Brown et al., 1989}), the hybrid Powell method of
MINPACK (\citeproc{ref-powell}{Powell, 1970}), and the
Runge--Kutta--Chebyshev (RKC) integration method
(\citeproc{ref-rkc}{Sommeijer et al., 1998}). The code was modernized
where possible, with many \texttt{go\ to} statements removed and
additional logic added to support our applications (see for example the
discussion on VODE in \citeproc{ref-castro_simple_sdc}{Zingale et al.,
2022}). We also make use of the C++ autodiff library
(\citeproc{ref-autodiff}{Leal, 2018}) to compute thermodynamic
derivatives required in the Jacobians of our reaction networks.

Microphysics uses header-only implementations of all functionality as
much as possible to allow for easier compiler inlining, which is
especially important in GPU kernels. We also leverage C++17
\texttt{if\ constexpr} templating to compile out unnecessary
computations for performance. Generally, the physics routines and
solvers are written to work on a single zone from a simulation code, and
in AMReX, a C++ lambda-capturing approach is used to loop over zones
(and offload to GPUs if desired). When used with an application code,
this design permits the simulation state data to be allocated directly
in GPU memory and left there for the entire simulation, with all physics
run directly on the GPU. Since each zone in a simulation usually will
have a different thermodynamic state, the integration of reaction
networks can lead to thread divergence issues. To help mitigate this
issue, we can cap the number of integration steps and either retry an
integration on a zone-by-zone basis with different tolerances or
Jacobian approximations or pass the failure back to the application code
to deal with. This strategy has been successful for many large scale
simulations (\citeproc{ref-Zingale_2025}{Zingale et al., 2025}).

Another key design feature is the separation of the reaction network
from the integrator. This allows us to easily experiment with different
integration methods (such as the RKC integrator) and also support
different modes of coupling reactions to a simulation code, including
operator splitting and spectral deferred corrections (SDC; see, e.g.,
\citeproc{ref-castro_simple_sdc}{Zingale et al., 2022}). The latter is
especially important for explosive astrophysical flows. Tight
integration with pynucastro (\citeproc{ref-pynucastro2}{Smith et al.,
2023}; \citeproc{ref-pynucastro}{Willcox \& Zingale, 2018}) allows for
the generation of custom reaction networks for a science problem.

There are two ways to use Microphysics: in a standalone fashion (via the
unit tests) for simple investigations or as part of an (AMReX-based)
application code. In both cases, the core (compile-time) requirement is
to select a network---this defines the composition that is then used by
most of the other physics routines. This compile-time requirement also
allows Microphysics to provide the number of species as a
\texttt{constexpr} value (which many application codes need), and
greatly reduces the compilation time (due to the templating used
throughout the library).

\section{Research impact statement}\label{research-impact-statement}

Microphysics has been used for simulations of convective Urca
(\citeproc{ref-Boyd_2025}{Boyd et al., 2025}) and X-ray bursts
(\citeproc{ref-Guichandut_2024}{Guichandut et al., 2024}) with
MAESTROeX; and for simulations of nova (\citeproc{ref-Smith2025}{Smith
Clark \& Zingale, 2025}), X-ray bursts
(\citeproc{ref-Harpole_2021}{Harpole et al., 2021}), thermonuclear
supernovae (\citeproc{ref-Zingale_2024_dd}{Zingale, Chen, Rasmussen, et
al., 2024}), and convection in massive stars
(\citeproc{ref-Zingale_2024}{Zingale, Chen, Johnson, et al., 2024}) with
Castro. This Microphysics library has also enabled recent work in
astrophysical machine learning to train deep neural networks modeling
nuclear reactions (\citeproc{ref-nn_astro_2022}{Fan et al., 2022};
\citeproc{ref-dnn_astro_2025}{X. Zhang et al., 2025}).

\section{AI usage disclosure}\label{ai-usage-disclosure}

No generative AI/LLM was used for producing code or documentation in the
git repository or for this paper. We have experimented with using AI/LLM
tools for code review and for suggesting places to focus our
optimization efforts on, but the resulting coding, benchmarking, and
testing is then done by humans.

\section{Acknowledgements}\label{acknowledgements}

The AMReX-Astro Microphysics library developers are an open scientific
team with members contributing to various aspects of the library. We
have thus chosen to display the members of our development team in the
author list in alphabetical order. All developers who have contributed
new features, substantial design input, and/or at least 3 commits were
invited to be coauthors. The work at Stony Brook was supported by the US
Department of Energy, Office of Nuclear Physics grant DE-FG02-87ER40317.

\section*{References}\label{references}
\addcontentsline{toc}{section}{References}

\protect\phantomsection\label{refs}

\end{document}